\documentstyle[lcws2002,psfig,epsfig,endnotes,longtable,pifont,boxedminipage,rotating,axodraw]{article}

\def\s#1{{\small#1}}

\def\be{\begin{equation}}
\def\ee{\end{equation}}
\def\bea{\begin{eqnarray}}
\def\eea{\end{eqnarray}}
\def\lsim{\:\raisebox{-0.5ex}{$\stackrel{\textstyle<}{\sim}$}\:}
\def\gsim{\:\raisebox{-0.5ex}{$\stackrel{\textstyle>}{\sim}$}\:}
\def\etal{ {\em et al.}}

\begin{document}

\begin{flushright}
{CERN-TH/2002-240\\
IPPP/02/55\\
DCPT/02/110\\
September 2002}
\end{flushright}

\title{
DETECTION OF HEAVY CHARGED HIGGS BOSONS AT FUTURE LINEAR COLLIDERS
VIA $\tau^-\bar\nu_\tau H^+$ PRODUCTION
%Detection of heavy charged Higgs bosons
%at future Linear Colliders\\
%via $\tau^-\bar\nu_\tau H^+$ production
}

\author{Stefano Moretti}
\address{CERN Theory Division, CH-1211 Geneva 23, Switzerland
and\\
Institute for Particle Physics Phenomenology, Durham DH1 3LE, UK}

\maketitle\abstracts{We show how a statistically significant signal 
of heavy charged Higgs bosons
of a Type II Two-Higgs Doublet (2HDM)
Model produced in association with tau-neutrino pairs 
can be established at future $e^+e^-$ Linear Colliders (LCs)
in the $H^+\to t\bar b\to 4$ jet decay channel for large $\tan\beta$
in the $M_{H^\pm}\gsim \sqrt s/2$ mass region.}

\vspace*{-0.75cm}

\section{Introduction}

Charged Higgs bosons, $H^\pm$,
 appear in the particle spectrum of a general Type II 2HDM, 
including the minimal Supersymmetric version of it:
the Minimal Supersymmetric Standard Model (MSSM). In
this context, the importance of singly produced charged Higgs bosons
at future LCs
 \cite{eebiblio} has been emphasised lately in several instances
 \cite{singleHpm,tbHpm,rest}. In fact, while 
the detection of $H^\pm$ states would represent an
unequivocal evidence of physics beyond the Standard Model (SM),
their mass could well be very large. For example, within the MSSM
in the so-called `decoupling-limit', one expects the following configuration
of masses among the five Higgs states of the model: $M_h\lsim130$ GeV
$\ll M_H\sim M_A\sim M_{H^\pm}$, for any $\tan\beta$. More over, if
only a light Higgs state is found at the Large Hadron Collider (LHC), 
the initial task of a LC  would be to 
start running at a rather low energy (say, $\sqrt s=350$ to 500 GeV), 
where the corresponding Higgs 
production cross section (via $e^+e^-\to Z^*\to Zh$) is largest. At such 
energies, the heavier Higgs states may not be produced in the leading  
channels  $e^+e^-\to ZH, A H, H^-H^+$
 \cite{Higgs,eeHpHm}, either because below threshold (i.e., $M_A+M_H, M_{H^+}+
M_{H^-}> \sqrt s$) or since the intervening MSSM coupling in the decoupling 
limit becomes zero (e.g., in the $ZZH$ vertex). Whereas in the neutral Higgs
sector the heavy $H$ and $A$ resonances can always be accessed via
$\gamma\gamma\to$
`triangle loop' $\to H/A$, this is not possible for the charged
Higgs boson states. Besides, in
the large $\tan\beta$ region, for neutral Higgs
states, one could 
alternatively resort to the associate production mode $e^+e^-\to b\bar b
H/A$. The corresponding channel for a charged Higgs boson would be
$e^+e^-\to b\bar t H^+$, which has an additional large mass in the final
state (i.e., $m_t=175$ GeV).  

\section{Single $H^\pm$ production}

Hence, it becomes clear the importance of studying
production modes of charged Higgs 
bosons with only one such particles in 
the final state.  An analysis of various single production modes
was performed in Ref.~\cite{singleHpm}, limitedly to their inclusive
rates. There, it was shown that
only two channels offer some chances of detection for
$M_{H^\pm}\gsim \sqrt s/2$, when $\sqrt s=500$ GeV \footnote{When $\sqrt s
\gg M_{H^\pm}+m_t$, the channel $e^+e^-\to b\bar t H^+$ becomes important
too \cite{tbHpm}. Furthermore, see Ref.~\cite{Shinya} for an 
illustration of the scope
of the $ \tau^-\bar\nu_\tau H^+$ final state in $\gamma\gamma$ 
collisions.}:
\begin{eqnarray}
e^+e^- &\to& \tau^-\bar\nu_\tau H^+, \tau^+\nu_\tau H^-~~
\mathrm{(tree\ level)},
\label{proc_tau} \\
e^+e^- &\to& W^\mp H^\pm ~~\mathrm{(one\ loop)}.    \label{proc_wh}
\end{eqnarray}
The former is relevant in the large $\tan\beta$ region, whereas the
latter is important for the low one. As LEP2 data seem to prefer large values 
of $\tan\beta$, at least in the MSSM \cite{LEPTRE}, the
first of these two processes has been analysed already against the 
background in  Ref.~\cite{taunuHpm}. We summarise the
finding of that paper, where a successful selection procedure
of channel (\ref{proc_tau}) was devised. A similar study for the case
of process (\ref{proc_wh}) is also in progress \cite{progress}.

The poor production rates  
of process (\ref{proc_tau}) reported in Ref.~\cite{singleHpm} require one 
to resort to the main decay channel of heavy charged Higgs
bosons, i.e., $H^+\to t\bar b$ \cite{BRs}, so that the signal ($S$)  and
(main) irreducible background ($B$) are:
\begin{eqnarray}\label{signal}
e^+ e^- &\rightarrow& \tau^- \bar\nu_\tau H^+ \to 
\tau^- \bar\nu_\tau t\bar b,\\ 
\label{background}
e^+ e^- &\rightarrow& \bar t t\to \tau^- \bar\nu_\tau t\bar b.
\end{eqnarray}  
One then  requires the emerging top to decay 
hadronically, i.e., $t\to bW^+\to
3$ jets, whereas one assumes the $\tau$'s  
to be tagged as narrow jets in their 
`one-prong' hadronic decays:
\begin{eqnarray}\nonumber
\tau^\pm &\rightarrow& \pi^\pm \nu_\tau~~~~~~~~~~~~~~~(12\%), \\ \nonumber
\tau^\pm &\rightarrow& \rho^\pm(\to \pi^\pm\pi^0) \nu_\tau~~(26\%),\\ \nonumber
\tau^\pm &\rightarrow& a_1^\pm (\pi^\pm\pi^0\pi^0)\nu_\tau~~~(8\%).
\end{eqnarray}
Altogether, the complete signature is: 
\begin{equation}\label{signature}
\tau-{\rm{jet}}~+~p_T^{\mathrm {miss}}~+~4~{\mathrm{jets}}.
\end{equation}

\section{Numerical results}

The simulation of Ref.~\cite{taunuHpm} was carried out at parton level,
for a Type II 2HDM, with $\tan\beta=40$  as reference value. Therein
the programs of Refs.~\cite{singleHpm,bbVV} were used, supplemented
by a finite calorimeter resolution
emulated through a Gaussian smearing of the jet  transverse momenta. 
A double tagging of $b$-jets in the final state was assumed throughout.  
 
\begin{figure}
\begin{center}
\epsfig{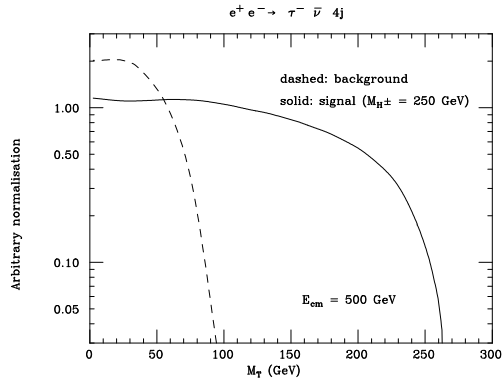}
\end{center}
\vspace{-0.5cm}
\caption{\small Distribution in the transverse mass (\ref{MT})
for processes (\ref{signal})--(\ref{background}) before cuts.}
\label{fig:MT}
\end{figure}

The cuts used were as follows. Like in Ref.~\cite{Battaglia},
the Cambridge jet clustering algorithm \cite{CAMBRIDGE}
was enforced to isolate a five jet sample, by using $y_{\rm cut}=0.001$,
wherein the $\tau$-jet was treated on the same footing as the quark-jets.
Besides, both $\tau$- and quark-jets were required to pass the following
cuts in energy and polar angle (hereafter, $j$ represents a generic jet):
\begin{equation} \label{Ethetacuts}
 E_{j} >  5    ~{\mathrm GeV},\qquad 
 |\cos\theta_{j}| <  0.995.
\end{equation}
Further assuming that the former can be distinguished from the latter
thanks to very different sub-jet distributions, one can apply a sequential 
$W^\pm$ and $t$ mass reconstruction only to quark-jets:
\begin{equation}\label{masscuts}
 |M_{jj}-M_{W^\pm}| <  10    ~{\mathrm GeV},\qquad 
 |M_{jjj}-m_t| <  15    ~{\mathrm GeV}.
\end{equation}
The cut in missing transverse momentum was:
\begin{equation}\label{pTmisscut}
p_T^{\rm miss} >  40    ~{\mathrm GeV}.
\end{equation}
A minimum transverse mass constructed from the 
visible $\tau$-jet and the missing transverse momentum
was also required:
\begin{equation}\label{MT} 
M_T \equiv\sqrt{2 p_T^{\tau}
{p}_T^{\mathrm{miss}} (1 - \cos\Delta\phi)}>M_{W^\pm},
\end{equation}
where $\Delta\phi$ is the relative azimuthal angle. Finally, 
following the findings of Refs.~\cite{dptau,tautau} in case of 
$H^\pm$ hadroproduction, it was imposed the requirement 
\begin{equation}\label{fraccut}
R_\tau={p^{\pi^\pm}}/{p_T^{\tau}}> 0.8,
\end{equation}
where $p^{\pi^\pm}$ is the momentum of the leading pion coming
from the $\tau$-lepton, and $p_T^{\tau}$ is the visible momentum
of the latter.

\begin{figure}
\begin{center}
\epsfig{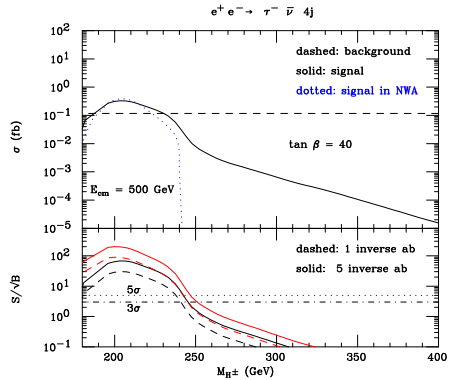}
\end{center}
\vspace{-0.5cm}
\caption{\small (Top) Total cross sections for processes 
(\ref{signal})--(\ref{background}) yielding the signature (\ref{signature}),
after the kinematic cuts in (\ref{Ethetacuts})--(\ref{fraccut}), including 
all decay BRs. For comparison, we also include the result for the signal
in Narrow Width Approximation (NWA) (blue).
(Bottom) Statistical significances of the signal
for two values of integrated
luminosity (the $3\sigma$ and $5\sigma$ `evidence' and `discovery' 
threshold are also given) after the kinematic
cuts in (\ref{Ethetacuts})--(\ref{fraccut}) (in black) and the
additional one in (\ref{M4jcut}) (in red).}
\label{fig:cuts}
\end{figure}

Fig.~\ref{fig:MT} illustrates the strong impact of the constraint in
transverse mass, by comparing the shape of the signal and background
before the kinematic selection. The signal distributions are obtained
at the points $M_{H^\pm}\approx \sqrt s/2$.

The upper plots in Fig.~\ref{fig:cuts} present the signal rates after the 
full kinematic selection has been enforced. (The background 
cross section is constant with $M_{H^\pm}$ as the above 
cuts do not depend on this
parameter.)  In the lower plots we display from Ref.~\cite{taunuHpm}
the significances
(in black) of the signal rates, after 1 and 5 ab$^{-1}$ of accumulated
luminosity, $\cal L$. 
It is clear that at this point neither evidence ($\gsim3\sigma$)
nor discovery ($\gsim5\sigma$) of charged Higgs bosons is possible
in the region $M_{H^\pm}\gsim\sqrt s/2$, whereas for
$M_{H^\pm}\lsim\sqrt s/2$ the signal should be comfortably observed.
The dominant contributions to the latter in this mass regimes come
from pair production $e^+e^-\to H^-H^+$, followed by
$H^-\to\tau^-\bar\nu_\tau$ decays. 

\begin{figure}
\begin{center}
\epsfig{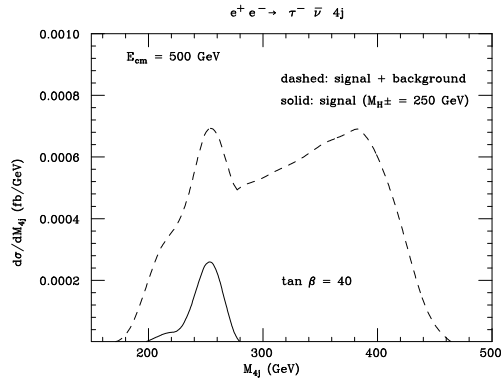}
\end{center}
\vspace{-0.5cm}
\caption{\small Distribution in the invariant mass of the four quark-jets
recoiling against the $\tau$-jet
for the sum of processes (\ref{signal})--(\ref{background}) 
and for the former separately yielding the signature (\ref{signature}),
after the
kinematic cuts in (\ref{Ethetacuts})--(\ref{fraccut}), including 
all decay BRs.}
\label{fig:mass}
\end{figure}

One can however exploit the invariant mass of the four-quark 
jet system. Fig.~\ref{fig:mass} shows this quantity. For the signal, it 
represents the reconstructed resonance of the charged Higgs boson (which was
originally generated at the point $M_{H^\pm}\approx \sqrt s/2$).
For the background, it correspond to a non-resonant kinematic
distribution. The width of the signal spectra is dominated by
detector smearing effects and suggests that a further selection
criterium can be enforced to enhance the $S/B$ rates, e.g.:
\begin{equation}\label{M4jcut}
|M_{4j}-M_{H^\pm}|<35~{\rm GeV}, 
\end{equation}
where the value used for $M_{H^\pm}$ would be the central 
or fitted mass resonance of the region in $M_{4j}$ were an excess of the form 
seen in Fig.~\ref{fig:mass} will be established.

The two red lines in Fig.~\ref{fig:cuts} show the significances
of the charged Higgs boson signals in presence of the constraint
in (\ref{M4jcut}), alongside those in (\ref{Ethetacuts})--(\ref{fraccut}).
These prove that one can expect to extend the reach 
in $M_{H^\pm}$ obtained from pair production of charged Higgs bosons and 
decays by about 50 GeV or so around and above the $M_{H^\pm}\approx \sqrt s/2$
point (see the dotted blue line in the top
plot of Fig.~\ref{fig:cuts}). 
Finally, recalling that $H^\pm$ rates have been given at 
$\tan\beta=40$ and that process (\ref{proc_tau}) is
proportional to the square of $\tan\beta$ for large values of the latter, 
the number of events in the threshold region would scale like 
$6(30)({\tan\beta}/{40})^2$, 
in correspondence of $\cal L=$ 1(5) ab$^{-1}$. 
(According to Ref.~\cite{Battaglia}, the single
$b$-tag efficiency is expected to be close to the value $\epsilon_b=90\%$, 
so that our main conclusions should remain unchanged even in presence of a
finite double-$b$-tagging efficiency.) An analysis of the considered
signature (\ref{signature}) within the full event generator environment
provided by \s{HERWIG} \cite{HERWIG} is also in progress \cite{MC}.

\vskip0.25cm
I thank the LCWS2002 organisers 
for the excellent atmosphere and stimulating environment that they 
have created during the workshop and The Royal Society of London, UK,
for partial financial support in the form of a Conference Grant.

\end{document}